\documentclass[12pt, letterpaper]{article}%
\usepackage{epsfig,amssymb,amsmath}        
\usepackage{algorithmic}
\usepackage{algorithm}
\usepackage{subfigure}
\newcounter{theorem}
\newtheorem{theorem}{\noindent\textbf{\emph{Theorem}}}

\newcounter{definition}
\newtheorem{definition}{\textbf{\emph{Definition}}}

\newcounter{claim} 

\newcounter{lemma} \newtheorem{lemma}{\noindent\textbf{\emph{Lemma}}}
\newcounter{proposition}

\renewcommand{\AA}{\mathcal A}

\newcommand{\FF}{\mathcal F}

\def\mapright#1{\smash{
   \mathop{\longrightarrow}\limits^{\tiny{#1}}}}
\begin{document}

\title{Near Optimal Routing in a Small-World Network \\with Augmented Local Awareness}

\author{Jianyang~Zeng\thanks{Center for Advanced Information Systems, Nanyang Technological University, Nanyang
Avenue, Singapore 639798. Email:\sf{ zengjy321@pmail.ntu.edu.sg.
}} \and Wen-Jing~Hsu\thanks{Nanyang Technological University,
Nanyang Avenue, Singapore 639798. Email:\sf{ hsu@ntu.edu.sg. }}
\and Jiangdian~Wang\thanks{School of Electrical Electronic
Engineering, Nanyang Technological University, Singapore.
Email:\sf{ wang0059@pmail.ntu.edu.sg. } } }

\setlength{\baselineskip}{16pt plus 1pt minus 1pt}

\bibliographystyle{plain}

\maketitle

\vspace{-10pt}
\begin{abstract}
In order to investigate the routing aspects of small-world
networks, Kleinberg~\cite{Kle00} proposes a network model based on
a $d$-dimensional lattice with long-range links chosen at random
according to the $d$-harmonic distribution. Kleinberg shows that
the greedy routing algorithm by using only local information
performs in $O(\lg^2 n)$ expected number of hops, where $n$
denotes the number of nodes in the network. Martel and
Nguyen~\cite{MN04analyzing} have found that the expected diameter
of Kleinberg's small-world networks is $\Theta(\lg n)$. Thus a
question arises naturally: Can we improve the routing algorithms
to match the diameter of the networks while keeping the amount of
information stored on each node as small as possible?\\

Existing approaches for improving the routing performance in the
small-world networks include: (1) Increasing the number of
long-range links~\cite{ADS02fault,MBR03symphony}; (2) Exploring
more nodes before making routing decisions~\cite{LS04almost};  (3)
Increasing the local awareness for each
node~\cite{FGP04eclecticism,MN04analyzing}. However, all these
approaches can only achieve $O\big( (\lg n)^{1+\epsilon} \big)$
expected number of hops, where $\epsilon >0$ denotes a constant.
We extend Kleinberg's model and add two augmented local links for
each node, which are connected to nodes chosen randomly and
uniformly within $\lg^2 n$ Mahattan distance. Our investigation
shows that these augmented local connections can make small-world
networks more
navigable.\\

We show that if each node is aware of $O(\lg n)$ number of
neighbors via the augmented local links, there exist both
non-oblivious and oblivious algorithms that can route messages
between any pair of nodes in $O(\lg n \lg \lg n)$ expected number
of hops, which is a near optimal routing complexity and
outperforms the other related results for routing in Kleinberg's
small-world networks. Our schemes keep only $O(\lg^2 n)$ bits of
routing information on each node, thus they are scalable with the
network size. Our results imply that the awareness of $O(\lg n)$
nodes through augmented links is more efficient for routing
than via the local links~\cite{FGP04eclecticism,MN04analyzing}.\\

Besides adding new light to the studies of social networks, our
results may also find applications in the design of large-scale
distributed networks, such as peer-to-peer systems, in the same
spirit of Symphony~\cite{MBR03symphony}.


\end{abstract}
\vspace{-10pt}
 \vspace{-10pt}
\section{Introduction}
A well-known study by Milgram in 1967~\cite{Mil67the} shows the
\emph{small-world phenomenon}~\cite{Koc89the}, also called ``six
degree of separation", that any two people in the world can be
connected by a chain of six (on the average) acquaintances, and
people can deliver messages efficiently to an unknown target via
their acquaintances. This study is repeated by Dodds, Muhamad, and
Watts~\cite{DMW03an} recently, and the results show that it is
still true for today's social network. The small-world phenomenon
has also been shown to be pervasive in networks from nature and
engineering systems, such as the World Wide
Web~\cite{WS98collective,AJB99the}, peer-to-peer
systems~\cite{ADS02fault,MNW04know,MBR03symphony,ZGG02using}, etc.

Recently, a number of network models have been proposed to study
the small-world
properties~\cite{New00models,WS98collective,Kle00}. Watts and
Strogatz~\cite{WS98collective} propose a random rewiring model
whose diameter is a poly-logarithmic function of the size of the
network. The model is constructed by adding a small number of
random edges to nodes uniformly distributed on a ring, where nodes
are connected densely with their near neighbors. A similar
approach can also be found in Ballab\'{a}s and Chung's earlier
work~\cite{BC88the}, where the poly-logarithmic diameter of the
random graph is achieved by adding a random matching to the nodes
of a cycle. However, these models fail to capture the algorithmic
aspects of a small-world network~\cite{Kle00}. As commented by
Kleinberg in~\cite{Kle00}, the poly-logarithmic diameter of some
graphs  does not imply the existence of efficient routing
algorithms. For example, the random graph in~\cite{BC88the} yields
a logarithmic diameter, yet any routing using only local
information requires at least $\sqrt{n}$ expected number of hops
(where $n$ is the size of the network)~\cite{Kle00}.

In order to incorporate routing or navigating  properties into
random graph models, Kleinberg~\cite{Kle00} develops a new model
based on a $d$-dimensional torus lattice with long-range links
chosen randomly from the $d$-harmonic distribution, i.e.,  a
long-range link between nodes $u$ and $v$ exists with probability
proportional to $Dist(u,v)^{-d}$, where $Dist(u,v)$ denotes the
Mahattan distance between nodes $u$ and $v$. Based on this model,
Kleinberg then shows that routing messages between any two nodes
can be achieved in $O(\lg ^2 n)$~\footnote{The logarithmic symbol
$\lg$ is with the base $2$, if not otherwise specified. Also, we
remove the ceiling or floor for simplicity throughout the paper.}
expected number of hops by applying a simple greedy routing
algorithm using only local information. This bound is tightened to
$\Theta(\lg^2 n)$ later by Barri\`{e}re et al.~\cite{BFK01} and
Martel et al.~\cite{MN04analyzing}. Further
research~\cite{MNW04know,LS04almost,MN04analyzing,FGP04eclecticism}
shows that in fact the $O(\lg^2 n)$ bound of the original greedy
routing algorithm can be improved by putting some extra
information in each message holder. Manku, Naor, and
Wieder~\cite{MNW04know} show that if each message holder at a
routing step takes its own neighbors' neighbors into account for
making routing decisions, the bound of routing complexity can be
improved to $O(\frac{\lg^2 n}{q\lg q})$, where $q$ denotes the
number of long-range contacts for each node. Lebhar and
Schabanel~\cite{LS04almost} propose a routing algorithm for
1-dimensional Kleinberg's model, which visits $O(\frac{\lg^2
n}{\lg^2 (1+q)})$ nodes on expectation before routing the message,
and they show that a routing path with expected length of
$O(\frac{\lg n(\lg \lg n)^2}{\lg^2(1+q)})$ can be found. Two
research groups, Fraigniaud et al.~\cite{FGP04eclecticism}, and
Martel and Nguyen~\cite{MN04analyzing}, independently report that
if each node is aware of its $O(\lg n)$ closest local neighbors,
the routing complexity in $d$-dimensional Kleinberg's small-world
networks can be improved to $O(\lg n \lg^{1+1/d}n)$ expected
number of hops. The difference is that~\cite{MN04analyzing}
requires keeping additional state information,
while~\cite{FGP04eclecticism} uses an oblivious greedy routing
algorithm. Fraigniaud et al.~\cite{FGP04eclecticism} also show
that $\Theta(\lg^2 n)$ bits of topological awareness per node is
optimal for their oblivious routing scheme.
In~\cite{MN04analyzing}, Martel and Nguyen show that the expected
diameter of a $d$-dimensional Kleinberg network is $\Theta(\lg
n)$. As such, there is still some room for reducing the routing
complexity, which motivates our work.

Other small-world models have also been studied. In their recent
paper~\cite{NM05analyzing}, Nguyen and Martel study the diameters
of variants of Kleinberg's small-world models, and provide a
general framework for constructing classes of small-world networks
with $\Theta(\lg n)$ expected diameter. Aspnes, Diamadi, and
Shah~\cite{ADS02fault} find that the greedy routing algorithms in
directed rings with a constant number of random extra links given
in \emph{any} distribution requires at least $\Omega(\lg^2 n/\lg
\lg n)$ expected number of hops. Another related models are the
small-world percolation
models~\cite{MNW04know,BB01diameter,CGS02diameter,Bis05submitted}.
The diameters of these models are studied by Benjamin et
al.~\cite{BB01diameter}, Coppersmith et al.~\cite{CGS02diameter}
and  Biskup~\cite{Bis05submitted}. The routing aspects of the
percolation models, such as the lower bound and upper bound of
greedy routing algorithms with 1-lookahead, are studied
in~\cite{MNW04know}.


Applications of small-world phenomenon in computer networks
include efficient lookup in peer-to-peer
systems~\cite{MNW04know,ADS02fault,MBR03symphony,ZGG02using},
gossip protocol in a communication network~\cite{KKD01spatial},
flooding routing in ad-hoc networks~\cite{HI03small}, and the
study of diameter of World Wide Web~\cite{AJB99the}, etc.

\subsection{Our Contributions}\vspace{-6pt}
We extend Kleinberg's structures of small-world models with slight
change. Besides having long-range and local links on the grid
lattice, each node is augmented with two extra links connected to
nodes chosen randomly and uniformly within $\lg ^2 n$ Mahattan
distance. Based on this extended model, we present near optimal
algorithms for decentralized routing with $O(\lg n)$ augmented
awareness. We show that if each node is aware of $O(\lg n)$ number
of nodes via the augmented neighborhood, there exist both
non-oblivious and oblivious routing algorithms that perform in
$O(\lg n \lg \lg n)$ expected number of hops (see
Theorem~\ref{theorem:non-oblivious} and
 Theorem~\ref{theorem:oblivious routing}).  Our investigation
constructively show that the augmented local connections can make
small-world networks more navigable.

A comparison of our algorithm with the other existing schemes is
shown in Table~\ref{table:comparisons}. Our decentralized routing
algorithms assume that each node can compute a shortest path among
a poly-logarithmic number of known nodes. Such an assumption is
reasonable since each node in a computer network is normally a
processor and can carry out such a simple computation. Our schemes
keep $O(\lg^2 n)$ bits of routing information stored on each node,
thus they are scalable with the increase of network size. Our
investigation shows that the awareness of $O(\lg n)$ nodes through
the augmented links is more efficient for routing than via the
local links~\cite{FGP04eclecticism,MN04analyzing}.

We note that besides adding new light to the studies of social
networks such as Milgram's experiment~\cite{Mil67the}, our results
may also find applications in the design of large-scale
distributed networks, such as peer-to-peer systems, in the same
spirit of Symphony~\cite{MBR03symphony}. Since the links in our
extended model are randomly constructed according to the
probabilistic distribution, the network may be less vulnerable to
adversarial attacks, and thus provide good fault tolerance.
\vspace{-10pt}
\begin{table}[t]
\begin{scriptsize}
  \centering
    \begin{tabular}{|c|c|c|c|}
    \hline
 Scheme & $\#$bits of awareness  & $\#$steps expected & Oblivious   \\
& & &  or Non-oblivious?\\
   \hline
    Kleinberg's greedy~\cite{Kle00,ADS02fault,MBR03symphony}& $O(q \lg n)$ & $O(\lg^2 n/q)$ & Oblivious \\
   \hline
 NoN-greedy~\cite{MNW04know}& $O(q^2 \lg n)$ & $O(\lg^2 n/(q\lg q))$ & Non-oblivious \\
    \hline
Decentralized algorithm in~\cite{LS04almost} & $O\big( \lg^2 n/ \lg (1+q) \big)$  & $O\big( (\lg n)^2/ \lg^2(1+q) \big)$ & Non-oblivious \\
 \hline
Decentralized algorithm~\cite{MN04analyzing} & $O(\lg^2 n)$ & $O\big( (\lg n)^{1+1/d}  \big)$ &  Non-oblivious  \\
\hline

Indirect-greedy algorithm~\cite{FGP04eclecticism} & $O(\lg^2 n)$ & $O\big( (\lg n)^{1+1/d}  \big)$ &  Oblivious  \\
\hline

Our algorithms for the & $O( \lg^2 n )$  & $O(\lg n \ \lg \lg n)$ & Both are provided\\
model with augmented awareness &  &  &    \\
\hline

 \end{tabular} \vspace{1pt}
\caption{\scriptsize{Comparisons of our decentralized routing
algorithms with the other existing schemes. In the first three
schemes
(in~\cite{Kle00,ADS02fault,MBR03symphony,MNW04know,LS04almost}),
we suppose that each node has $q$ long-range contacts, while in
the next three schemes (in~\cite{MN04analyzing,FGP04eclecticism}
and this paper), we suppose that each node has one long-range
contact. A routing protocol is \emph{oblivious} if the message
holder makes routing decisions only by its local information and
the target node, and independently of the previous routing
history, otherwise, it is said to be \emph{non-oblivious}. }
  }\label{table:comparisons}
\end{scriptsize}
 \end{table}


\subsection{Organization}
The rest of the paper is organized as follows.
Section~\ref{sec:Preliminaries} gives  notations for Kleinberg's
small-world model and its extended version with augmented local
connections. Section~\ref{sec:decentralized routing} gives some
preliminary notations for decentralized routing. In
Section~\ref{sec:suboptimal routing}, we propose both
non-oblivious and oblivious routing algorithms with near optimal
routing complexity in our extended model.
Section~\ref{sec:Experimental} gives the experimental evaluation
of our schemes. Section~\ref{sec:conclusion} briefly concludes the
paper.


\vspace{-6pt}
\section{Definitions of Small-World Models} \label{sec:Preliminaries}\vspace{-6pt}
In this section, we will give the definition of Kleinberg's
small-world model and its extended version in which each node has
extra links. For simplicity, we only consider the one-dimensional
model with \emph{one} long-range contact for each node. In
addition, we assume that all links are directed, which is
consistent with the real-world observation, for example, person
$x$ knows person $y$, but $y$ may not know $x$. \vspace{-5pt}
\begin{definition} \emph{\textbf{(Kleinberg's Small-World Network
(KSWN)~\cite{Kle00})}} A Kleinberg's Small-World Network, denoted
as $\mathcal{K}$, is based on a one-dimensional torus (or ring)
$[n]=[0,1, \cdot \cdot \cdot, n]$. Each node $u$ has a directed
local link to its next neighbor $(u+1) \mod n$ on the ring. We
refer to this local link as \textbf{Ring-link} (or \textbf{R-link}
for short), and refer to node $(u+1) \mod n$ as the
 \textbf{R-neighbor} of node
$u$. In addition, each node has \emph{one} long-range link to
another node chosen randomly according to the 1-harmonic
distribution, that is, the probability that node $u$ sends a
long-range link to node $v$ is $\Pr[u \rightarrow v]=\frac{1}{Z_u
\cdot {Dist(u,v)}}$, where $Dist(u,v)$ denotes the ring
distance~\footnote{or Mahattan distance for multi-dimensional
models.} from $u$ to $v$, and $Z_u=\sum_{z\neq
u}\frac{1}{{Dist(u,z)}}$. We refer to this long-range link as the
\textbf{Kleinberg-link} (or \textbf{K-link} for short), and refer
to node $v$ as a \textbf{K-neighbor} of node $u$ if a K-link
exists from $u$ to $v$.
\end{definition}\vspace{-5pt}

Our extended structure introduces several extra links for each
node. Its definition is given below.

\vspace{-5pt}
\begin{definition}
\emph{\textbf{(KSWN with Augmented Local Connections (KSWN*))}} A
Kleinberg's Small-World Network with Augmented Local Connections,
denoted as $\mathcal{K^*}$, has the same structure of KSWN, except
that each node $u$ in $\mathcal{K^*}$ has two extra links to nodes
chosen randomly and uniformly from the interval  $(u,u+\lg^2 n]$.
We refer to these two links as the \textbf{augmented local links}
(or \textbf{AL-links} for short), and refer to node $v$ as a
\textbf{AL-neighbor} of node $u$ if a AL-link exists from $u$ to
$v$.
\end{definition}
\vspace{-5pt}

There are in total four links for each node in a KSWN*: one
R-link, one K-link, two AL-links. We refer to all nodes linked
directly by node $u$ as the \textbf{immediate neighbors} of $u$.
Our extended structure retains the same $O(1)$ order of node
degree as that of Kleinberg's original model.


\vspace{-6pt}
\section{Decentralized Routing Algorithms}\label{sec:decentralized routing}\vspace{-6pt}
Based on the original model, Kleinberg presents a class of
decentralized routing algorithms, in which each node makes routing
decisions by using local information and in a greedy fashion. In
other words, the message holder forward the message to its
immediate neighboring node, including its K-neighbor, which is
closest to the destination in terms of the Mahattan distance.
Kleinberg shows that such a simple greedy algorithm performs in
$O(\lg ^2 n)$ expected number of hops. The other existing
decentralized routing
algorithms~\cite{ADS02fault,MBR03symphony,LS04almost,FGP04eclecticism,MN04analyzing,MNW04know}
mainly rely on three approaches to improve routing performance:
(1) Increasing the number of long-range
links~\cite{ADS02fault,MBR03symphony}; (2) Exploring more nodes
before making routing decisions~\cite{LS04almost};  (3) Increasing
the local awareness for each
node~\cite{FGP04eclecticism,MN04analyzing}. However, so far using
these approaches can only achieve $O\big( (\lg n)^{1+\epsilon}
\big)$ expected number of hops in routing, where $\epsilon>0$.
Although the scheme in~\cite{MNW04know}, where each node makes
routing decision by looking ahead its neighbors's neighbors, can
achieve an optimal $O(\lg n/\lg \lg n )$ bound, their result
depends on the fact that each node has at least $\Omega(\lg n)$
number of K-links.

There are normally two approaches for decentralized routing:
oblivious and non-oblivious schemes~\cite{FGP04eclecticism}.  A
routing protocol is \emph{oblivious} if the message holder makes
routing decisions only by its local information and the target
node, and independently of the previous routing history. On the
other hand, if the message holder needs to consider certain
information of the previous routing history to make routing
decisions, the protocol is referred to as \emph{non-oblivious}.
The non-oblivious protocol is often implemented by adding a header
segment to the message packet so that the downstream nodes can
learn the routing decisions of upstream nodes by reading the
message header information. The scheme in~\cite{FGP04eclecticism}
is oblivious, while the schemes in~\cite{LS04almost}
and~\cite{MN04analyzing} are non-oblivious.

We refer to the message holder as \emph{the current node}. For the
current node $x$, we define a sequence of node sets $T_0, T_1,
\cdot \cdot \cdot , T_i, \cdot \cdot \cdot$, where $T_0=\{ x \}$,
$T_1=\{$ $u$'s AL-neighbors, $\forall u \in T_0 \}$, $T_2=\{
\mbox{$u$'s AL-neighbors}, \forall u \in T_1 \}$, and so on. We
refer to $T_i$ as the set of nodes in \emph{the $i$th level of AL
neighborhood}, and let $H_i=\bigcup_{j\leq i} T_j$ denote the set
of all nodes in \emph{the first $i$ levels of AL neighborhood}. At
a certain level $i$ of AL neighborhood, we may also refer to
$H_{i-1}$ as the set of \emph{previously known nodes}. Let
$L_i=T_i-H_{i-1}$ denote the set of \emph{new} nodes discovered
during the $i$th level of AL neighborhood. Let $A_x (k)=H_{k}$
denote the \emph{augmented local awareness} (or \emph{AL
awareness} for short) of a given node in a KSWN*, where each node
is aware of the first $k$ levels of its AL neighborhood.

In Section~\ref{sec:suboptimal routing}, we will show that there
exists a sufficiently large constant $\sigma$ such that $|A_x(\lg
\lg n)|\geq \lg n/\sigma$, based on which we propose both
non-oblivious and oblivious routing algorithms running in $O(\lg n
\lg \lg n)$ expected number of hops and requiring $O(\lg ^2 n)$
bits of information on each node.

Our near optimal $O(\lg n \lg \lg n)$ bound on the routing
complexity outperforms the other related results for Kleinberg's
small-world networks. To our knowledge, our algorithms achieve the
best expected routing complexity while requiring at most $O(\log
^2 n)$ bits of information stored on each node.


\vspace{-6pt}
\section{Near Optimal Routing with $O(\lg
n)$ Awareness}\label{sec:suboptimal routing} \vspace{-6pt}

\subsection{Augmented Local Awareness of $O(\lg n)$}\vspace{-6pt}
In this subsection, we will show that $|A_x(\lg \lg n)|$, the
number of distinct nodes that node $x$ is aware of via the first
$\lg \lg n$ levels of AL neighborhood, is not less than $ \lg n/
\sigma$ for a constant $\sigma$, which, as will be shown in
Lemma~\ref{lemma:half distance suboptimal}, is sufficiently large
to guarantee that $A_x (\lg \lg n)$ contains a K-link that jumps
over half distance (Suppose that the destination node is at a
certain large distance from the current node). These results are
useful for the subsequent analysis of our oblivious and
non-oblivious routing schemes.

\vspace{-6pt}
\begin{lemma} \label{lemma:aware_logn}
Let $A_x (\lg \lg n)$ denote the AL awareness of node $x$ in a
KSWN* $\mathcal{K^*}$, where each node  is aware of $\lg \lg n$
levels of AL-neighbors. Then
$$\Pr
[\ | A_x (\lg \lg n)|\geq \frac{\lg n}{\sigma} \ ]>\psi,$$ where
$\sigma$ denotes a sufficiently large constant and $\psi$ denotes
a positive constant.

\end{lemma}
\vspace{-8pt} \noindent \textbf{Proof:}\quad Throughout the proof,
we assume that $|H_i|< \frac{\lg n}{\sigma}$ for all $1\leq i \leq
\lg \lg n$, otherwise, the lemma already holds, since $|A_x(\lg
\lg n)|=|H_{\lg \lg n}|>\lg n/\sigma$. We will show that at each
level of AL neighborhood, the probability that each AL-link points
to previously known nodes is small so that a large number of
distinct nodes will be found via the first $\lg \lg n$ levels of
AL neighborhood.

Consider the construction of a AL-link for the current node $x$.
By definition of KSWN*, each AL-link of $x$ is connected to a node
randomly and uniformly chosen from the interval $(x,x+\lg^2 n]$,
that is, each AL-link of $x$ points to a node in the interval
$(x,x+\lg^2 n]$ with probability $(\lg n)^{-2}$. By assumption,
there could be no more than $\lg n/ \sigma$ previously known nodes
in the interval $(x,x+\lg^2 n]$. Thus, the probability for a
AL-link of a given node to point to a previously known node is at
most $(\lg n/\sigma) \cdot (\lg n)^{-2}=(\sigma \lg n)^{-1}$.
Thus, the probability for a AL-link of $x$ to point to a new node
is at least $1-(\sigma \lg n)^{-1}$. There are in total at most $2
\cdot |H_{\lg \lg n}| \leq 2 \lg n/\sigma$ number of AL-links, so
the probability for all AL-links to point to new nodes is at least
$(1-(\sigma \lg n)^{-1} )^{2 \lg n/\sigma}\geq 1-
\frac{2}{\sigma^2 }$ for sufficiently large $n$. Here we use the
fact $(1+x)^a \geq 1+ax$ for $x>-1$ and $a\geq 1$. When $\sigma$
is a sufficiently large constant, we have $\Pr [\ |A_x|\geq
\frac{\lg n}{\sigma}\ ]> \psi$ for a positive constant $\psi=1-
\frac{2}{\sigma^2 }>0$. Thus, the proof of
Lemma~\ref{lemma:aware_logn} is completed.~\
\hfill\rule{2.0mm}{2.6mm}

\vspace{-6pt}
\subsection{Non-Oblivious Decentralized Routing} \label{subsec:suboptimal nonoblivious}\vspace{-6pt}
Our non-oblivious routing algorithm is given as follows: Initially
the source node $s$ finds in its AL awareness $A_s(\lg \lg n)$ an
intermediate node $z$ that is closest to the destination, and then
computes a shortest path $\pi$ from $s$ to $z$ in $A_s(\lg \lg
n)$. Before routing the message, $s$ adds the information about
shortest path $\pi$ to the message header. Once the message passes
a node on the shortest path $\pi$, the next stop is read off the
header stack. When the message reaches node $z$, node $z$ can tell
that it is an intermediate target by reading the message header
and then route the message to its K-neighbor. Such processes are
repeated until the message reaches a certain node close enough to
the destination node. After that, Kleinberg's plain greedy
algorithm can be used to route the message effectively to the
target node. Given a message $M$, a source node $s$ and a target
node $t$ in a KSWN* $\mathcal{K^*}$, the pseudocodes of our
non-oblivious algorithm running on the current node $x$ are given
in Algorithm~\ref{alg:non-oblivious}.

\begin{algorithm*}
\begin{scriptsize}
\caption{} \label{alg:non-oblivious}
\begin{algorithmic}
\item[\textbf{Input:}] the source $s$, the target $t$ and the
message $M$.
\end{algorithmic}

\begin{algorithmic}
\item[\textbf{Initialization:}]   \STATE   $\mbox{Current node}
\leftarrow s$.
 \STATE Set the header stack of the message $M$ to be
empty.
\end{algorithmic}

\begin{algorithmic}
\item[\textbf{while} Distance between the current node and the
destination $\geq (\lg n)^2 \lg \lg n$ \textbf{do}] \IF{the header
stack of the message $M$ is empty} \STATE Route the message $M$ to
$x$'s K-neighbor $y$. \STATE Find an intermediate node $z$ in
$A_y(\lg \lg n)$ whose K-neighbor is closest to $t$ (ties are
broken arbitrarily). \STATE Compute a shortest path $\pi:x_0=y,
x_1, \cdot \cdot \cdot, x_t=z$ from $y$ to $z$, and push the
shortest path information $\pi: x_1, \cdot \cdot \cdot, x_t=z$
into the header stack of the message $M$. \ELSE
 \STATE Pop up the first node $x_i$ from the header stack and
route the message $M$ to node $x_i$.    \ENDIF

\item[\textbf{end while}]
\end{algorithmic}

\begin{algorithmic}
\item[\textbf{Final phase (Kleinberg's greedy algorithm):}] \STATE
Route the message $M$ to an immediate neighbor of $x$ that is
closest to the target $t$, until it reaches $t$.
\end{algorithmic}

\end{scriptsize}\vspace{-3pt}
\end{algorithm*}

Next we will analyze the performance of the
Algorithm~\ref{alg:non-oblivious}. We first give a basic lemma,
which provide  a lower bound and an upper bound on the probability
of the existence of a K-link in Kleinberg's small-world networks.
Its proof can be found in Appendix A.

\vspace{-2pt}
\begin{lemma} \label{lemma:lower_upper_probablity}
Let $\Pr [u \mapright{K} v]$ denote the probability that node $u$
sends a K-link to node $v$ in a KSWN* $\mathcal{K^*}$. Suppose
that $a \leq {Dist(u,v)} \leq b$, then $\frac{c_1}{b \lg n}\leq
\Pr [u \mapright{K} v] \leq \frac{c_2}{a \lg n}$, where $c_1$ and
$c_2$ are constants independent of $n$.
\end{lemma}

In Lemma~\ref{lemma:aware_logn}, we have shown that $\Pr[\ |A_x
(\lg \lg n)|\geq \lg n/\sigma\ ]$ is at least a positive constant
for a sufficiently large constant $\sigma$. Based on this result,
Lemma~\ref{lemma:half distance suboptimal} shows that the
probability for $A_x(\lg \lg n)$ to contain a K-link jumping over
half distance is at least a positive constant.

\vspace{-5pt}
\begin{lemma} \label{lemma:half distance suboptimal}
Suppose that the distance between the current node $x$ and the
target node $t$ in a KSWN* $\mathcal{K^*}$ is $Dist(x,t) \geq
\lg^2 n \lg \lg n$. Then with probability at least a positive
constant, node $x$'s AL awareness $A_x(\lg \lg n)$ contains a
K-neighbor within $Dist(x,t)/2$ distance to the target node $t$ .
\end{lemma}
\vspace{-5pt}

\noindent \textbf{Proof:}\quad Let $\AA$ denote the event that
$|A_x(\lg \lg n)|\geq \frac{\lg n}{\sigma}$. By
Lemma~\ref{lemma:aware_logn}, we have $\Pr [\AA]>\psi$ for a
constant $\psi>0$.

Let $B_l(t)$ denote the set of all nodes within $l$ ring distance
to $t$. Let $\Pr[x \mapright{K} B_l(t)]$ denote the probability
that $x$'s K-neighbor is inside the ball $B_l(t)$.

Let $m=Dist(x,t)$. By Lemma~\ref{lemma:lower_upper_probablity},
the probability for a K-link to point to a given node inside the
ball $B_{\frac{m}{2}}(t)$ is at least $\frac{c_1}{m \lg n}$, so we
have  \vspace{-6pt}
\[
\Pr[x \mapright{K} B_{\frac{m}{2}}(t)] \geq
|B_{\frac{m}{2}}(t)|\cdot \frac{c_1 }{m \lg n} =\frac{m}{2} \cdot
\frac{c_1 }{m \lg n}   \geq \frac{c_3 }{\lg n},
\]\vspace{-2pt}
where $c_3$ is a constant.

Since $Dist(x,t) \geq \lg^2 n \lg \lg n$ and each AL-link spans a
distance no more than $\lg^2 n$, the nodes in AL awareness
$A_x(\lg \lg n)$ are all between the current node $x$ and the
target node $t$. Let $\Pr[A_x (\lg \lg n) \mapright{K}
B_{\frac{m}{2}}(t)]$ denote the probability that at least one node
in $A_x(\lg \lg n)$ has a K-neighbor in $B_{\frac{m}{2}}(t)$. Then
we have

\vspace{-8pt}
\begin{eqnarray*}
 \Pr[A_x (\lg \lg n) \mapright{K} B_{\frac{m}{2}}(t)]&& \geq \Pr[A_x (\lg \lg n)\mapright{K}
B_{\frac{m}{2}}(t)\mid \AA] \cdot \Pr [\AA] \\
&& \geq \big(1-(1-\frac{c_3}{\lg n})^{\frac{\lg n}{\sigma}}
\big)\cdot \psi \\
&&\geq \psi (1-e^{-\frac{c_3}{\sigma}} ),
\end{eqnarray*}\vspace{-11pt}

which is larger than a positive constant. At the last step, we
obtain $(1-\frac{c_3}{\lg n})^{\frac{\lg n}{\sigma}}\leq
e^{-\frac{c_3}{\sigma}}$  by using the fact that
$(1+\frac{b}{x})^x\leq e^b$ for $b\in \mathbb{R}$ and $x>0$.~\
\hfill\rule{2.0mm}{2.6mm}

\vspace{-6pt}
\begin{lemma} \label{lemma:large distance routing for suboptimal}
Suppose that the distance between the current node $x$ and the
target node $t$ in a KSWN* $\mathcal{K^*}$ is $Dist(x,t) \geq
\lg^2 n \lg \lg n$. Then after at most $O(\lg n \lg \lg n)$
expected number of hops, Algorithm~\ref{alg:non-oblivious} will
reduce the distance to within $\lg^2 n \lg \lg n$.
\end{lemma}\vspace{-4pt}

\noindent \textbf{Proof:}\quad Since $Dist(x,t)\geq \lg^2 n \lg
\lg n$, all known nodes in $x$'s AL awareness $A_x(\lg \lg n )$
are between the current node $x$ and the target node $t$. We can
apply the result in Lemma~\ref{lemma:half distance suboptimal} to
analyze Algorithm~\ref{alg:non-oblivious}.

We refer to the routing steps from a given node $x$ to any node
within $A_x(\lg \lg n)$ as an indirect phase. The routings in
different indirect phases are independent from each other. By
Lemma~\ref{lemma:half distance suboptimal}, the probability that
node $x$'s AL awareness $A_x(\lg \lg n)$ contains a K-neighbor
within $Dist(x,t)/2$ distance to the target node $t$ is at least a
positive constant, so after at most $O(1)$ expected number of
indirect phases, Algorithm~\ref{alg:non-oblivious} will find an
intermediate node whose K-link jumps over half distance. Since
each indirect phase takes at most $\lg \lg n$ hops and the maximum
distance is $n$, after at most $O(\lg n \ \lg \lg n)$ expected
number of hops, the message will reach a node within $\lg^2 n \lg
\lg n$ distance to the target node $t$.~\
\hfill\rule{2.0mm}{2.6mm}

\vspace{-5pt}
\begin{lemma} \label{lemma:last phase for suboptimal}
Suppose that the distance between the current node $x$ and the
target node $t$ in a *KSWN $\mathcal{^*K}$ is $Dist(x,t) \leq
\lg^2 n \lg \lg n$. Then using the final phase of
Algorithm~\ref{alg:non-oblivious} (i.e. using Kleinberg's greedy
algorithm) can route the message to the target node $t$ in $O(\lg
n)$ expected number of hops.
\end{lemma}\vspace{-5pt}

\noindent \textbf{Proof:}\quad When the distance $Dist(x,t) \leq
\lg^2 n \lg \lg n$, the final phase in
Algorithm~\ref{alg:non-oblivious} is executed. By Kleinberg's
results in~\cite{Kle00}, after at most $O\big( \lg^2(  \lg^2 n \lg
\lg n ) \big)=O(\log n)$ expected number of steps, the message
will be routed to the destination node.~\
\hfill\rule{2.0mm}{2.6mm}

Combining the above lemmas, it is not difficult for us to obtain
the routing complexity of Algorithm~\ref{alg:non-oblivious}.
\vspace{-5pt}
\begin{theorem} \label{theorem:non-oblivious}
In a KSWN* $\mathcal{K^*}$, Algorithm~\ref{alg:non-oblivious}
performs in $O(\lg n \ \lg \lg n)$ expected number of hops.
\end{theorem}\vspace{-4pt}

\vspace{-6pt}
\subsection{Oblivious Decentralized Routing} \label{subsec:suboptimal oblivious}\vspace{-6pt}
In our oblivious scheme, when the distance is large, the current
node $x$ first finds in $A_x(\lg \lg n)$ whether there is an
intermediate node $z$, which contains a K-neighbor within $Dist
(x,t)/2 $ distance to the target node, and is closest to node $x$
in terms of AL-links (any possible tie is broken arbitrarily).
Next, node $x$ computes a shortest path $\pi$ from $x$ to $z$
among the AL awareness $A_x(\lg \lg n)$, and then routes the
message to its next AL-neighbor on the shortest path $\pi$. When
the distance is small, Kleinberg's plain greedy algorithm is
applied.

Given a message $M$, a source $s$ and a target $t$ in a KSWN*
$\mathcal{K^*}$, the pseudocodes of our oblivious algorithm
running on the current node $x$ are given in
Algorithm~\ref{alg:oblivious}.
\vspace{-8pt}
\begin{algorithm}
\begin{scriptsize}
\caption{} \label{alg:oblivious}

\begin{algorithmic}
\item[\textbf{Input:}] the source $s$, the target $t$ and the
message $M$.
\end{algorithmic}

\begin{algorithmic}
\item[\textbf{Initialization:}] \STATE   $\mbox{Current node}
\leftarrow s$.
\end{algorithmic}

\begin{algorithmic}
\item[\textbf{while} Distance between the current node and the
destination $\geq c(\lg n)^2 \lg \lg n$  \textbf{do}]  \ \ \ \
($c$ is a sufficiently large constant and will be given later)

\STATE $z \leftarrow$ a node in $A_x(\lg \lg n)$ that contains a
K-neighbor within $Dist(x,t)/ 2$ distance to $t$, and is closest
to node $x$ in terms of AL-links (ties are broken arbitrarily).

\IF{node $z$ does not exist} \STATE Route the message $M$ to an
immediate neighbor closest to node $t$. \ELSE \STATE Compute a
shortest path $\pi$ from $x$ to $z$ among $A_x(\lg \lg n)$.
\IF{$\pi$ consists of only node $x$ itself} \STATE Route the
message $M$ to the K-neighbor. \ELSE \STATE Route the message $M$
to the next AL-neighbor on the shortest path $\pi$. \ENDIF \ENDIF

\item[\textbf{end while}]
\end{algorithmic}

\begin{algorithmic}
\item[\textbf{Final phase (Kleinberg's greedy algorithm):}] \STATE
Route the message $M$ to an immediate neighbor of $x$ that is
closest to the target $t$, until it reaches $t$.
\end{algorithmic}

\end{scriptsize}\vspace{-3pt}
\end{algorithm}
\vspace{-4pt}

\vspace{-5pt}
\begin{lemma} \label{lemma:half_large}
Suppose that the distance between the current node $x$ and the
target node $t$ in a KSWN* $\mathcal{K^*}$ is $Dist(x,t) \geq c
(\lg n)^2 \lg \lg n$, where $c$ is a sufficiently large constant.
Then after at most $O(\lg \lg n)$ expected number of hops,
Algorithm~\ref{alg:oblivious} will reduce the distance to within
$Dist(x,t)/2$.
\end{lemma}\vspace{-4pt}

\begin{figure*}
\centering
\includegraphics[width=4in]{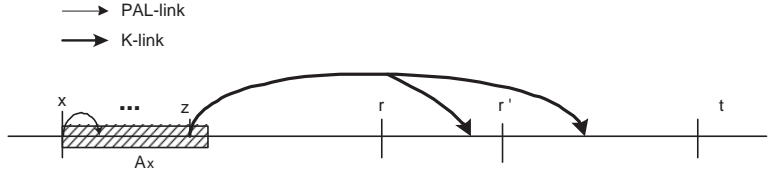}
\caption{\scriptsize{Diagram for oblivious decentralized routing.
The shade area represents node $x$'s AL awareness $A_x(\lg \lg
n)$. The target node $t$ is on the right side of $x$. Node $r$ is
the midpoint of $\overline{xt}$. Node $r'$ is between nodes $r$
and $t$ such that $Dist(r,r')=\lg^2 n \ \lg \lg n$. Node $z$ is an
intermediate node in $A_x(\lg \lg n)$ that contains a K-neighbor
in $\overline{rt}$ (in $\overline{rr'}$ or $\overline{r't}$ ) and
is closest to $x$ in terms of AL-links. }} \label{fig:KSWN}
\end{figure*}

\noindent \textbf{Proof:}\quad As shown in Figure~\ref{fig:KSWN},
node $r$ is the midpoint of  $\overline{xt}$, and node $r'$ is
between $r$ and $t$ such that $Dist(r, r')=\lg^2 n \lg \lg n$. Let
$z$ be an intermediate node in $A_x(\lg \lg n)$ that contains a
K-neighbor between $r$ and $t$, and is closest to $x$ in terms of
AL-links. We refer to a node $z$ in $x$'s AL awareness $A_x(\lg
\lg n)$ as a \emph{good} intermediate node if it satisfies the
following two conditions: (1) has a K-neighbor within
$Dist(x,t)/2$ to the target node; (2) is closest to node $x$ in
terms of AL-links. Let $\pi: x_0=x, x_1, \cdot \cdot \cdot, x_t=z$
denote a shortest path that $x$ finds from itself to $z$ among the
AL awareness $A_x(\lg \lg n)$. We divide the next routing into two
cases according to the different locations of $z$'s K-neighbor.

\vspace{-6pt}
\vspace{-6pt}

In the first case, $z$'s K-neighbor is within $\overline{r' t}$.
Since the distance between $x$ and the right most node in $A_x(\lg
\lg n)$ is at most $\lg ^2 n \lg \lg n$, $z$'s K-neighbor is also
within $Dist(x_i,t)/2$ to the target node for every $x_i$ on the
shortest path $\pi$, that is, node $z$ always satisfies the first
condition of a good intermediate node for every node $x_i$. Also,
if $z$ is an intermediate node closest to $x$, it is also a
closest intermediate node to every $x_i$ on the shortest path
$\pi$, that is, $z$ also satisfies the second condition of a good
intermediate node for every node $x_i$. Therefore, node $z$ will
become a fixed good intermediate node for all nodes $x_i$ on the
shortest path. When this case happens,
Algorithm~\ref{alg:oblivious} will route the message along a
shortest path from $x$ to $z$ in an oblivious routing fashion.
Thus, in this case, after at most $\lg \lg n$ number of hops, the
message will reach a good intermediate node and the routing
distance will be reduced by half~\footnote{There may be more than
one good intermediate nodes $z$ when a tie happens. However, even
when this happens, the message will still reach one of good
intermediate nodes along a shortest path finally. Hereinafter, we
focus on the case in which the good intermediate node $z$ is
unique for the current node $x$. The analysis for the case with
multiple good intermediate nodes can be easily obtained. }. In the
second case, $z$'s K-neighbor is within $\overline{rr'}$. When
this happens, the intermediate node $z$ may change for each $x_i$
on the shortest path $\pi: x_0=x, x_1, \cdot \cdot \cdot, x_t=z$,
and the message may not be routed along the shortest path as
expected by the previous node $x$. However, we will show that the
latter case will not happen very likely, since the length of
$\overline{rr'}$ is relatively small when $Dist(x,t) \geq c (\lg
n)^2 \lg \lg n$ for a sufficiently large constant $c$.

Let $\FF_1$ denote the event that $A_x(\lg \lg n)$ contains a
K-neighbor in $\overline{r't}$. By using a similar technique in
Lemma~\ref{lemma:half distance suboptimal}, we can easily obtain
that $\FF _1$ occurs with probability at least a positive
constant.

Let $\FF_2$ denote the event that $A_x(\lg \lg n)$ contains a
K-neighbor in $\overline{rr'}$. For any node $y$ in $A_x(\lg \lg
n)$, we have $Dist(y,r)\geq \frac{1}{3}c (\lg n)^2 \lg \lg n$ when
$c$ is a sufficiently large constant. By
Lemma~\ref{lemma:lower_upper_probablity}, the probability for a
node $y$ in $A_x(\lg \lg n)$ to send a K-link to a node in
$\overline{rr'}$ is at most $\frac{3 c_2}{c (\lg n)^2 (\lg \lg n)
\cdot \lg n}$. Because there are in total $\lg^2 n \lg \lg n$
nodes in $\overline{rr'}$, a node in $A_x(\lg \lg n)$ has a
K-neighbor in $\overline{rr'}$ with probability at most $\frac{3
c_2}{c (\lg n)^2 (\lg\lg n) \cdot \lg n } \cdot \lg^2 n \lg \lg n
=\frac{3 c_2}{c \lg n}$. Since $|A_x(\lg \lg n)|\leq 1+2+2^2
+\cdot \cdot \cdot +2^{\lg \lg n} \leq 2\lg n$, the event $\FF_2$,
i.e., $A_x(\lg \lg n)$ has a K-neighbor in $\overline{rr'}$,
occurs with probability at most $\frac{3 c_2}{c \lg n} \cdot 2 \lg
n =\frac{6c_2}{c}$, which is smaller than a certain constant when
$c$ is a sufficiently large constant. Thus, we have $Pr[\neg
\FF_2]>\gamma$ for a constant $\gamma>0$, if we choose a
sufficiently large constant $c$.

Therefore, $\Pr[\neg{\FF_2}\bigcap \FF_1]$ is larger than a
positive constant, if we choose a sufficiently large constant $c$.
Thus, after at most $c' \lg \lg n$ expected number of hops for a
constant $c'$, the event $\neg \FF_2 \bigcap \FF_1$ will occur,
that is, a message will be routed to a node $x$ whose AL awareness
$A_x(\lg \lg n)$ contains a K-neighbor in $\overline{r' t}$, but
no K-neighbor in $\overline{rr'}$. When such a node $x$ is
reached, the intermediate node $z$ is fixed for every node $x_i$
on a shortest path $\pi:x_0=x, x_1, \cdot \cdot \cdot, x_t=z$ in
an oblivious routing fashion. Then after at most $\lg \lg n$
number of hops, the message will be routed to the fixed
intermediate node $z$, which has a K-link jumping over half
distance.

Therefore,  after at most $c' \lg \lg n + \lg  \lg n=O(\lg \lg n)$
expected number of hops, the distance will be reduced by half.~\
\hfill\rule{2.0mm}{2.6mm}

\vspace{-4pt}
\begin{lemma} \label{lemma:large distance oblivous routing for suboptimal}
Suppose that the distance between the current node $x$ and the
target node $t$ in a KSWN* $\mathcal{K^*}$ is $Dist(x,t) \geq
c\lg^2 n \lg \lg n$, where $c$ is a sufficiently large constant.
Then after at most $O(\lg n \lg \lg n)$ expected number of hops,
Algorithm~\ref{alg:oblivious} will reduce the distance to within
$c\lg^2 n \lg \lg n$.
\end{lemma}\vspace{-4pt}
\noindent \textbf{Proof:}\quad The proof is similar to that of
Lemma~\ref{lemma:large distance routing for suboptimal}, and hence
is omitted here.~\ \hfill\rule{2.0mm}{2.6mm}

\vspace{-4pt}
\begin{lemma} \label{lemma:half_small}
Suppose that the distance between current node $x$ and the target
node $t$ in a KSWN* $\mathcal{K^*}$ is $m< c (\lg n)^2 \lg \lg n$,
where $c$ is a sufficiently large constant. Then using the final
phase of Algorithm~\ref{alg:oblivious} (i.e. using Kleinberg's
greedy algorithm) can route the message to the target node $t$ in
$O(\lg n)$ expected number of hops.
\end{lemma}\vspace{-4pt}

\noindent \textbf{Proof:}\quad The proof is similar to that of
Lemma~\ref{lemma:last phase for suboptimal}, and hence is omitted
here.~\ \hfill\rule{2.0mm}{2.6mm}

Combining the above lemmas, we can easily obtain the following
theorem.

\vspace{-4pt}
\begin{theorem} \label{theorem:oblivious routing}
In a KSWN* $\mathcal{K^*}$, Algorithm~\ref{alg:oblivious} performs
in $O(\lg n \ \lg \lg n)$ expected number of hops.
\end{theorem}\vspace{-4pt}

\vspace{-6pt}
\section{Experimental Evaluation}\label{sec:Experimental}
In this section, we will conduct experiments to evaluate our
schemes and other existing routing schemes for Kleinberg's
small-world networks.

We focus on the following four schemes: (a) The original greedy
routing algorithm~\cite{Kle00} in Kleinberg's small-world network
with only one long-range contact per node. Each node forwards the
message to its immediate neighbor closest to the destination; (b)
The greedy routing algorithm in Kleiberg's small-world network
with two long-range contacts per
node~\cite{ADS02fault,MBR03symphony}. In the experimental study,
we would like to learn how much the additional number of
long-range links can help routing. (c) The decentralized routing
scheme with $O(\lg n)$ local
awareness~\cite{FGP04eclecticism,MN04analyzing}. With this scheme,
we intend to evaluate the degree at which the local awareness
improve the routing efficiency. (d) Our near optimal routing
scheme proposed in this paper.  We note that most schemes have
both non-oblivious and oblivious versions. Here we only focus on
the non-oblivious version for each scheme.

\subsection{Experimental Setup}

\noindent \textbf{Network Construction:} We construct the
small-world network based on a ring $[0, 1, \cdot \cdot \cdot,
n]$. Each node $i$ is connected to its immediate neighbors $(i+1)
\mod n$. Let $H_n= \sum_{i=1}^n 1/i$ denote the harmonic
normalization factor. We then generate a sequence of intervals
$I(i)$, which we call the probability intervals, where $1\leq i
\leq n-1$. Let $0< I_1 \leq 1/H_n$, and $\frac{1}{(i-1)H_n} <I_i
\leq \frac{1}{i H_n}$, where $2\leq i \leq n -1$. Each node $i$
uniformly generates a random number  $x$ in $(0,1]$, and then
finds the interval that contains $x$. Suppose that $x$ is located
in the interval $I_k$. Node $i$ then forms a long-range link
connected to a node with the distance $k$. When each node has
multiple long-range contacts, it just generates more than one
random numbers, and sets up the connections in the same way.

In the extension of Kleinberg's small-world networks, each node
uniformly and randomly chooses two nodes within the Manhattan
distance $\lg^2 n$ as its augmented local neighbors.

\noindent \textbf{Messages Generation and Evaluation Metrics}: We
let each node generate a query message with a random destination,
and then evaluate the following metrics.

\begin{enumerate}
\item[(1)] \textbf{Average length of routing path} is the average
number of hops travelled by the messages from the source to the
destination. \item[(2)] \textbf{Storage requirement} for each node
is the number of information bits required to be stored on each
node.
\end{enumerate}

\subsection{Experimental Results}
\begin{figure*}[h]
\centering
\includegraphics[width=4in]{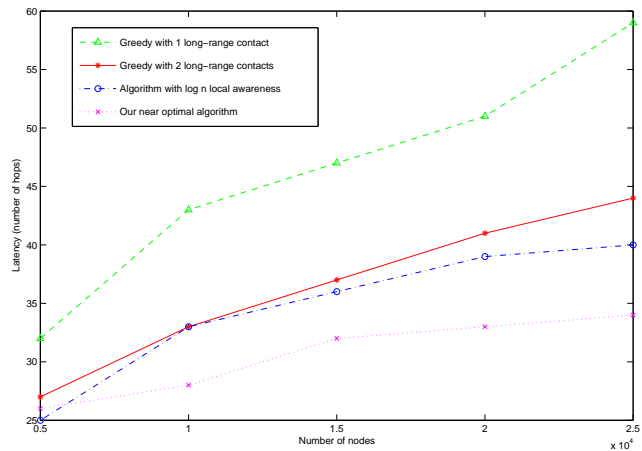}
\caption{Average length of routing paths for different routing
schemes.} \label{fig:static routing_small_world}
\end{figure*}

\begin{figure*}[h]
\centering
\includegraphics[width=4in]{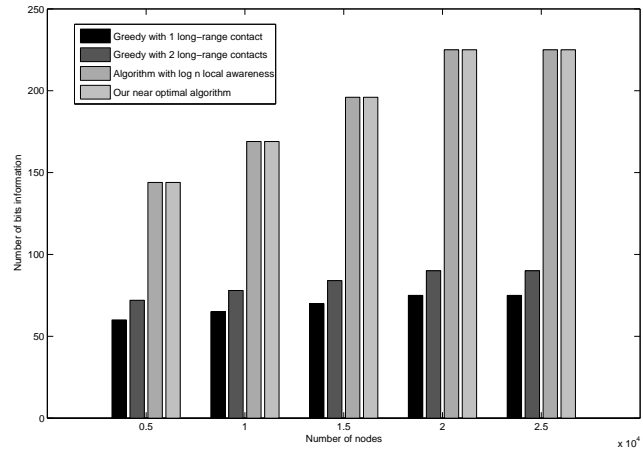}
\caption{Storage requirement on each node for different routing
schemes.} \label{fig:static_small_world_space}
\end{figure*}

We vary the number of nodes in the network from 5,000 to 25,000,
and evaluate different routing schemes, as shown in
Figures~\ref{fig:static routing_small_world} and
~\ref{fig:static_small_world_space}. For large $n$, the greedy
algorithm with increasing number of long-range
contacts~\cite{ADS02fault,MBR03symphony}, the decentralized
routing algorithm with local
awareness~\cite{FGP04eclecticism,MN04analyzing}, our near optimal
and algorithm all improve Kleinberg's original greedy algorithm.
Our near optimal scheme can find a shorter routing path than the
decentralized routing schemes with local
awareness~\cite{FGP04eclecticism,MN04analyzing}, while keep almost
the same storage space on each node.


\vspace{-6pt}
\section{Conclusion}\label{sec:conclusion}
We extend Kleinberg's small-world network with augmented local
links, and show that if each node participating in routing is
aware of $O(\lg n)$ neighbors via augmented links, there exist
both non-oblivious and oblivious decentralized algorithms that can
finish routing in $O(\lg n \lg \lg n)$ expected number of hops,
which is a near optimal routing complexity. Our investigation
shows that the awareness of $O(\lg n)$ nodes through the augmented
links will be more efficient for routing than via the local
links~\cite{FGP04eclecticism,MN04analyzing}.

Our extended model may provide an important supplement for the
modelling of small-world phenomenon, and may better approximate
the real-world observation. For example, each person in a human
society is very likely to increase his/her activities randomly
within some certain communities, and thus is aware of certain
levels of ``augmented" acquaintances. This augmented awareness
would surely help delivery the message to an unknown target in the
society.

Our results may also find applications in the design of
large-scale distributed networks, such as distributed storage
systems. Unlike most existing deterministic frameworks for
distributed systems, our extended small-world networks may provide
good fault tolerance, since the links in the networks are
constructed probabilistically and less vulnerable to adversarial
attacks.

%
%


%
%
%


\noindent \textbf{Appendix A. \ Proof of
Lemma~\ref{lemma:lower_upper_probablity} }\\

\noindent\textbf{Lemma 2.} \ \emph{Let $\Pr [u \mapright{K} v]$
denote the probability that node $u$ sends a K-link to node $v$ in
a KSWN* $\mathcal{K^*}$. Suppose that $a \leq {Dist(u,v)} \leq b$,
then $\frac{c_1}{b \lg n}\leq \Pr [u \mapright{K} v] \leq
\frac{c_2}{a \lg n}$, where $c_1$ and $c_2$ are constants
independent of $n$.}\\

\noindent \textbf{Proof:}\quad The probability that node $v$ is a
K-neighbor of node $u$ is $\Pr [u \mapright{K} v]
=\frac{1}{{Dist(u,v)} Z_v}$, where $Dist(u,v)$ is the ring
distance between nodes $u$ and $v$,  and $Z_v=\sum_{z \neq
v}\frac{1}{{Dist(v,z)}}$.
%

Observe that $Z_v=\sum_{i=1}^{n}\frac{|U_i|}{i}$, where $|U_i|$ is
the set of all nodes at distance $i$ away to node $v$. Since
$|U_i|=\Theta(1)$, we have $Z_v=\sum_{i=1}^{n}
\frac{\Theta(1)}{i}= \Theta(\lg n)$.

Since $a \leq {Dist(u,v)} \leq b$, we have $\frac{c_1}{{b} \lg n}
<\Pr [u \mapright{K} v] <\frac{c_2}{a \lg n}$, for some constants
$c_1$ and $c_2$ independent of $n$. Thus the lemma follows.~\
\hfill\rule{2.0mm}{2.6mm}\\ \\ \\ \\

%
%

\end{document}